\documentstyle[aps]{revtex}
\draft

\begin{document}
\author{Wei-Min Sun$^a$, Xiang-Song Chen$^{a,b}$ and Fan Wang$^a$ } 
\address{$^a$Department of Physics and Center for Theoretical Physics, 
	Nanjing University, Nanjing 210093, China\\
        $^b$Institut f\"ur Theoretische Physik, Universit\"at T\"ubingen,
                Auf der Morgenstelle 14, D-72076 T\"ubingen, Germany}
\title{Difficulties in a Kind of Averaging Procedure for Constructing Gauge-invariant Operators out of Gauge-variant Ones }
\maketitle
 
\begin{abstract}
We prove that a kind of averaging procedure for constructing gauge-invariant  
operators(or functionals) out of gauge-variant ones is erroneous and inapplicable  for a large class of operators(or functionals).

\end{abstract}

Gauge invariance is a basic requirement in gauge theories. Only gauge-invariant
operators correspond to physical observables. Then how to construct gauge-invariant operators(or functionals) out of gauge-variant ones is of fundamental importance since the advent of these theories. In the literature this problem
has been intensively studied. For example in \cite{Lavelle} a kind of dressing is used to construct gauge-invariant constituent quark and gluon fields out of gauge-variant bare ones. In the literature there also exist a kind of recipe of formally averaging out any gauge dependence to produce a gauge invariant  functional \cite{Kogan}. In this paper we will show that the latter method is erroneous and inapplicable for a large class of operators(or functionals). Here we restrict ourself in quantum electrodynamics.
 
First let us describe this averaging procedure. For an arbitrary functional $O[\phi]$, gauge invariant or not, we can always 
define a functional $G_O[\phi]=\int D\omega O[\phi^{\omega}]$ 
where $\phi$ denotes gauge or matter fields, $\phi^{\omega}$ denotes the 
result of $\phi$ after a gauge transformation $\omega$ and $\int D\omega$ 
stands for functional integration over the gauge group(here we choose the normalization so that $\int D\omega=1$). Using the property 
$\int D\omega f(\omega)=\int D\omega f(\omega_0\omega)$ we can easily prove 
that $G_O[\phi]$ is gauge-invariant:
\begin{eqnarray}
G_O[\phi^{\omega_0}] &=& \int D\omega O[\phi^{\omega_0\omega}]
 \nonumber \\
&=& \int D\omega O[\phi^{\omega}] \nonumber \\
&=& G_O[\phi]
\end{eqnarray}

This procedure formally gives a gauge-invariant functional. One nontrivial example is provided by the Faddeev-Popov trick \cite{Faddeev} in quantizing a gauge field theory using functional integral methods. The Faddeev-Popov determinant $\Delta_F[A]$ defined by
\begin{equation}
\Delta_F[A]\int D\omega \delta(F[A^{\omega}])=1 
\end{equation}
is gauge-invariant where $F[A]$ is a suitable gauge-fixing functional.
 
In the following we will show that for a class of functionals this kind of averaging procedure is erroneous and inapplicable.

First let us take $O[\phi]$ to be $A^{\mu}(x)$. In this case we will show that the averaging procedure is inapplicable. Now
$O[\phi^{\omega}]=A^{\mu}(x)+
\partial^{\mu}\theta(x)$ and formally integrating over $\omega$ yields
\begin{equation}
G_O[\phi]=A^{\mu}(x)+
\int D\omega \partial^{\mu}\theta(x)    
\end{equation}

From this we see that the averaging procedure gives a gauge-variant operator! Then where do we go wrong? The reason 
is the following: one can show that the functional integration 
$\int D\omega\partial^{\mu}\theta(x)$ cannot be consistently defined \cite{SCW}. Here we present the essential arguments. In any viable definition of functional integration over the gauge group the two properties $\int D\omega f(\omega)=\int D\omega f(\omega^{-1})$ and $\int D\omega f(\omega_0\omega)=\int D\omega f(\omega)$ must hold. Noting that $\partial^{\mu}\theta(x)=-i\omega^{-1}(x)\partial^{\mu}\omega(x)$, where $\omega(x)=e^{i\theta(x)}$, we have
\begin{equation}
\int D\omega \partial^{\mu}\theta(x) \sim \int D\omega \omega^{-1}(x)\partial^{\mu}\omega(x)
\end{equation}
Now we can show that the above mentioned two properties cannot simultaneously be satisfied when the integrand is $\omega^{-1}(x)\partial^{\mu}\omega(x)$. If we require the first property to hold this integration should be zero:
\begin{eqnarray}
\int D\omega \omega^{-1}(x)\partial^{\mu}\omega(x)&=&\int D\omega\omega(x)\partial^{\mu}\omega^{-1}(x) \nonumber \\
&=& -\int D\omega \omega^{-1}(x)\partial^{\mu}\omega(x)
\end{eqnarray}
If we require the second property to hold this integration should be infinity:
\begin{eqnarray}
\int D\omega\omega^{-1}(x)\partial^{\mu}\omega(x) &=& \int D\omega(\omega_0(x)\omega(x))^{-1}\partial^{\mu}(\omega_0(x)\omega(x)) \nonumber \\
&=& \int D\omega(\omega_0^{-1}(x)\partial^{\mu}\omega_0(x)+\omega^{-1}(x)\partial^{\mu}\omega(x)) \nonumber \\
&=& \omega_0^{-1}(x)\partial^{\mu}\omega_0(x)+\int D\omega\omega^{-1}(x)\partial^{\mu}\omega(x)
\end{eqnarray}
So from the above analysis we see that the functional integration $\int D\omega \partial^{\mu}\theta(x)$ cannot be consistently defined.
Now the situation is clear. The operator 
$\int D\omega O[\phi^{\omega}]$ does not exist so of course we cannot talk about 
its gauge invariance. Therefore in this case the averaging procedure is erroneous and inapplicable. 

In the following we will investigate a general functional of $A^{\mu}(x)$ of the form:
\begin{equation}
K[A;x]=K^{(0)}(x)+\int dy K^{(1)}_{\mu}(y;x)A^{\mu}(y)+\frac{1}{2!}\int\int dydzK^{(2)}_{\mu\nu}(y,z;x)A^{\mu}(y)A^{\nu}(z)+...
\end{equation}
where $K^{(0)}(x)$, $K^{(1)}_{\mu}(y;x)$, $K^{(2)}_{\mu\nu}(y,z;x)$ etc. are suitable smooth functions or distributions in their arguments. We also assume that $K^{(1)}$, $K^{(2)}$ etc. have compact supports with respect to the variables $y$, $z$ etc. Functionals expressible in this form should be called analytic functionals. 

Now we will derive a criteria for the gauge invariance of a general analytic functional. For this purpose we do an infinitesimal gauge transformation 
$(A^{\mu}(y))^{\theta}=A^{\mu}(y)+\partial^{\mu}\theta(y)$
and take $K[A;x]=K[A^{\theta};x]$. Expanding both sides in terms of $A^{\mu}(y)$ and comparing the zeroth order, first order $\cdots$ terms gives(here we only keep $\theta(y)$ up to first order because it is infinitesimal):
\begin{equation}
K^{(0)}(x)=K^{(0)}(x)+\int dyK^{(1)}_{\mu}(y;x)\partial^{\mu}\theta(y)
\end{equation}
\begin{equation}
\int dyK^{(1)}_{\mu}(y;x)A^{\mu}(y)
=\int dyK^{(1)}_{\mu}(y;x)A^{\mu}(y) 
+\frac{1}{2!}\int\int dy dz K^{(2)}_{\mu\nu}(y,z;x)(A^{\mu}(y)\partial^{\nu}\theta(z)+\partial^{\mu}\theta(y)A^{\nu}(z)) 
\end{equation}
~~~~~~~~~~~~~~~~~~~~~~~~~~~~~~~~~~~~~~~~~~~~~~~~$\cdots\cdots$

\noindent Using eq(8) and integrating by parts gives(this is always possible because $K^{(1)}_{\mu}(y;x)$ has compact support with respect to $y$)
\begin{equation}
\partial^{\mu}_{(y)}K^{(1)}_{\mu}(y;x)=0
\end{equation}
Similarly from eq(9) and using the symmetric property  $K^{(2)}_{\mu\nu}(y,z;x)=K^{(2)}_{\nu\mu}(z,y;x)$ we get
$\int dyK^{(2)}_{\mu\nu}(y,z;x)\partial^{\mu}\theta(y)=0$. 
 Using this equation and integrating by parts(for the same reason as above) gives
\begin{equation}
\partial^{\mu}_{(y)}K^{(2)}_{\mu\nu}(y,z;x)=0
\end{equation}
(Note that from the symmetric property of $K^{(2)}$ this equation automatically implies $\partial^{\nu}_{(z)}K^{(2)}_{\mu\nu}(y,z;x)=0$).
From this kind of reasoning we can obtain the criteria for the gauge invariance of an analytic functional $K[A;x]$
\begin{eqnarray}
\partial^{\mu}_{(y)}K^{(1)}_{\mu}(y;x)&=&0 \nonumber \\
\partial^{\mu}_{(y)}K^{(2)}_{\mu\nu}(y,z;x)&=&0 ~~~~~~\&~~~~~~ \partial^{\nu}_{(z)}K^{(2)}_{\mu\nu}(y,z;x)=0 \nonumber \\
~~~~~~~~~~~~~~~~~~~\cdots\cdots
\end{eqnarray}
From this criteria we can see that an analytic functional is gauge-invariant if and only if each term in its Taylor expansion is gauge-invariant separately.    

Now let us consider a polynomial-type analytic functional(an analytic functional with only finite terms in its Taylor expansion) and assume it is gauge-variant. It is possible that some terms of it is gauge-invariant. If this is the case we can drop these gauge-invariant pieces because they do not affect the averaging procedure. Then we get a gauge-variant polynomoial-type analytic functional with its highest order term in $A$ being gauge-variant. Now it is easy to see that the problem encountered in the averaging procedure for $A^{\mu}(x)$(the undefinability of the relevant functional integral) must also exist in this case. Because if it does not formally integrating with respect to $\omega$ would yield a polynomial-type analytic functional with the same highest order term as the original one, which must then be gauge-variant. That is to say, in this case the averaging procedure must be erroneous and inapplicable.

From the above analysis we can draw a conclusion: the averaging procedure is inapplicable for any gauge-variant local operator made out of the $A$ field and/or its derivatives. This is because any such local operator can be expressed as some polynomial-type analytic functional in $A$.(here is an examples: $(\partial\cdot A(x))^{2}=\int\int dy dz \partial^{(x)}_{\mu}\delta(y-x)\partial^{(x)}_{\nu}\delta(z-x)A^{\mu}(y)A^{\nu}(z)$).

Now we have proved that the averaging procedure is inapplicable for any gauge-variant polynomial-type analytic functional. In other words for the averaging procedure to be physically meaningful we must pay attention to analytic functional with infinite terms in its Taylor expansion \cite{Kogan} or non-analytic functionals(for example the case of Faddeev-Popov trick where a distribution in $A$ is involved).

In summary we have proved that the averaging procedure for constructing gauge-invariant functionals(or operators) out of gauge-variant ones must be inapplicable for a class of functionals(gauge-variant polynomial-type analytic functional in gauge fields)and especially for any gauge-variant local operator made out of gauge field and/or its derivatives. 
This work is supported in part by the NSF(19675018), SED and SSTD of China,
and in part by the DAAD.

\end{document}